\documentclass[aps,prb, twocolumn,superscriptaddress]{revtex4-2}
\usepackage{graphicx}
\usepackage[hidelinks]{hyperref} 
\usepackage{bm}
\usepackage{amsmath}
\usepackage{multirow}
\hypersetup{colorlinks,
}

\graphicspath{{./}}

\begin{document}

%\preprint{}

\title{Weakly coupled alternating $S=1/2$ chains in the distorted honeycomb lattice compound Na$_2$Cu$_2$TeO$_6$}
\thanks{This manuscript has been authored by UT-Battelle, LLC under Contract No. DE-AC05-00OR22725 with the U.S. Department of Energy.  The United States Government retains and the publisher, by accepting the article for publication, acknowledges that the United States Government retains a non-exclusive, paid-up, irrevocable, world-wide license to publish or reproduce the published form of this manuscript, or allow others to do so, for United States Government purposes.  The Department of Energy will provide public access to these results of federally sponsored research in accordance with the DOE Public Access Plan (http://energy.gov/downloads/doe-public-access-plan).}%

\renewcommand*{\thefootnote}{\arabic{footnote}}

\author{Shang Gao}
%\email[]{sgao.physics@gmail.com}
\affiliation{Materials Science \& Technology Division, Oak Ridge National Laboratory, Oak Ridge, TN 37831, USA}
\affiliation{Neutron Scattering Division, Oak Ridge National Laboratory, Oak Ridge, TN 37831, USA}

\author{Ling-Fang Lin}
%\email[]{sgao.physics@gmail.com}
\affiliation{Department of Physics and Astronomy, University of Tennessee, Knoxville, Tennessee 37996, USA }

\author{Andrew F. May}
%\email[]{sgao.physics@gmail.com}
\affiliation{Materials Science \& Technology Division, Oak Ridge National Laboratory, Oak Ridge, TN 37831, USA}

\author{Binod K. Rai}
%\email[]{sgao.physics@gmail.com}
\affiliation{Materials Science \& Technology Division, Oak Ridge National Laboratory, Oak Ridge, TN 37831, USA}

\author{Qiang Zhang}
%\email[]{sgao.physics@gmail.com}
\affiliation{Neutron Scattering Division, Oak Ridge National Laboratory, Oak Ridge, TN 37831, USA}

\author{Elbio Dagotto}
%\email[]{sgao.physics@gmail.com}
\affiliation{Materials Science \& Technology Division, Oak Ridge National Laboratory, Oak Ridge, TN 37831, USA}
\affiliation{Department of Physics and Astronomy, University of Tennessee, Knoxville, Tennessee 37996, USA }

\author{Andrew D. Christianson}
%\email[]{sgao.physics@gmail.com}
\affiliation{Materials Science \& Technology Division, Oak Ridge National Laboratory, Oak Ridge, TN 37831, USA}

\author{Matthew B. Stone}
%\email[]{sgao.physics@gmail.com}
\affiliation{Neutron Scattering Division, Oak Ridge National Laboratory, Oak Ridge, TN 37831, USA}


\date{\today}

%\keywords{}

\begin{abstract}
Spin-1/2 chains with alternating antiferromagnetic (AF) and ferromagnetic (FM) couplings exhibit quantum entanglement like the integer-spin Haldane chains and might be similarly utilized for quantum computations. Such alternating AF-FM chains have been proposed to be realized in the distorted honeycomb-lattice compound  Na$_2$Cu$_2$TeO$_6$, but to confirm this picture a comprehensive understanding of the exchange interactions including terms outside of the idealized model is required. Here we employ neutron scattering to study the spin dynamics in Na$_2$Cu$_2$TeO$_6$ and accurately determine the coupling strengths through the random phase approximation and density functional theory (DFT) approaches. We find the AF and FM intrachain couplings are the dominant terms in the spin Hamiltonian, while the interchain couplings are AF but perturbative. This hierarchy in the coupling strengths and the alternating signs of the intrachain couplings can be understood through their different exchange paths. Our results establish Na$_2$Cu$_2$TeO$_6$ as a weakly-coupled alternating AF-FM chain compound and reveal the robustness of the gapped ground state in alternating chains under weak interchain couplings.
\end{abstract}

\maketitle

\textit{Introduction.} Spin-1/2 chains with alternating AF and FM couplings are known to exhibit gapped excitations and exponentially decaying correlations~\cite{hida_ground_1992, hida_excitation_1994}, which are different from those of the spin-1/2 Bethe chains with uniform couplings~\cite{bethe_zur_1931} but more similar to the spectrum of integer-spin Haldane chains~\cite{haldane_nonlinear_1983, affleck_rigorous_1987}. Assuming alternating couplings of $J_1$ (AF) and $J_2 = \beta J_1$ ($\beta<1$), the  Hamiltonian of an alternating chain can be written as $\mathcal{H} = \sum_j J_1\bm{S}_{2j-1}\cdot \bm{S}_{2j} +  \sum_j J_2\bm{S}_{2j}\cdot \bm{S}_{2j+1}$. In the special case of $\beta = 0$, an alternating chain can be viewed as disconnected spin dimers, and the local singlet-triplet (triplon) excitations over the dimers account for the excitation gap in the spectrum. Non-zero $J_2$ couplings will introduce dispersion for the triplon excitations. At  $\beta = 1$, one recovers the gapless character of a Bethe chain. In the whole range of $-\infty<\beta<1$, theoretical calculations have revealed a hidden string order that is protected by the $\rm{Z_2}\times \rm{Z_2}$ global rotation symmetry~\cite{hida_crossover_1992, kohmoto_hidden_1992}, showing that the gapped ground state of alternating chains is a symmetry-protected topological state of the same type as that of Haldane chains~\cite{pollmann_entanglement_2010}. Considering the prospect of Haldane chains as the resource ground state for measurement-based quantum computation~\cite{miyake_quantum_2010}, the additional degrees of freedom in $S=1/2$ alternating chains may be important to explore further flexibility in qubit operations.
%---------------------------------------------------------------------
\begin{figure}[t!]
\includegraphics[width=0.46\textwidth]{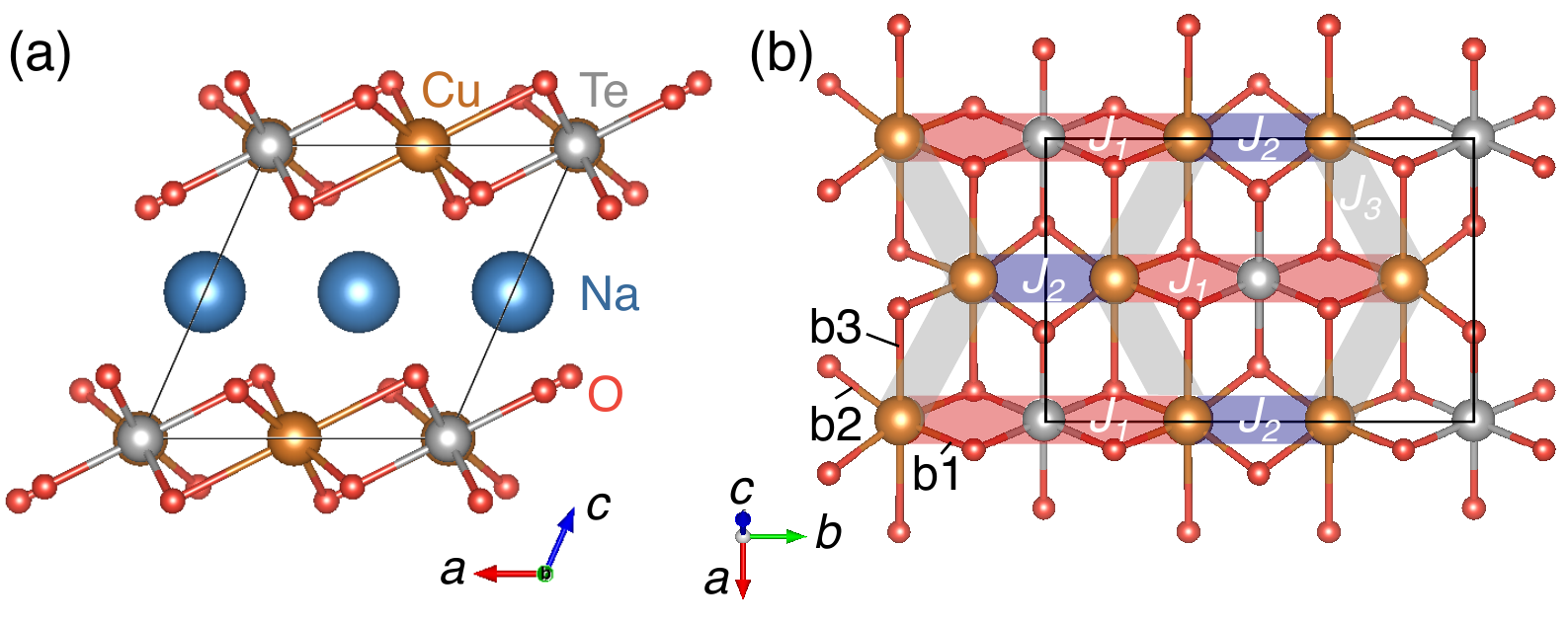}
\caption{(a, b) Crystal structure of Na$_2$Cu$_2$TeO$_6$ viewed along the $b$ (a) and $c^*$ (b) axes with the Cu$^{2+}$ ions forming chains along the $b$ axis. The size of the unit cell is indicated by black lines.  The Cu$^{2+}$ spins are coupled through alternating $J_1$ (red) and $J_2$ (blue) interactions along the chain and $J_3$ interactions (grey) between the chains. Cu-O bonds of different lengths are indicated in the CuO$_6$ octahedron  at the left bottom corner of panel (b), with b1 = 1.978(1) \AA, b2 = 1.999(1) \AA, and b3 = 2.533(1) \AA . 
\label{fig:xrd}}
\end{figure}
%---------------------------------------------------------------------

Experimental realizations of alternating AF-FM  chains are limited to a few compounds, including CuNb$_2$O$_6$~\cite{kodama_neutron_1999}, DMACuCl$_3$~\cite{stone_quantum_2007}, Na$_3$Cu$_2$SbO$_6$~\cite{miura_magnetic_2008}, and BaCu$_2$V$_2$O$_8$~\cite{klyushina_hamiltonian_2018} where the exchange couplings have been accurately determined through neutron scattering. Recently, a new candidate compound, Na$_2$Cu$_2$TeO$_6$, was proposed~\cite{xu_synthesis_2005, miura_spin_2006, derakhshan_electronic_2007, koo_analysis_2008, schmitt_microscopic_2014}. Similar to Na$_3$Cu$_2$SbO$_6$, the Cu$^{2+}$ ions ($S=1/2$) in Na$_2$Cu$_2$TeO$_6$ form a distorted honeycomb lattice in the $ab$ plane, which are separated by the Na$^{+}$ layers along the $c$ axis (see Fig.~\ref{fig:xrd}). Magnetic susceptibility measurements on a powder sample of Na$_2$Cu$_2$TeO$_6$ revealed a spin gap $\Delta\sim127$~K~\cite{xu_synthesis_2005}, which has been attributed to strong AF couplings $J_1$ in the DFT calculations~\cite{xu_synthesis_2005, derakhshan_electronic_2007, koo_analysis_2008, schmitt_microscopic_2014}. However, controversy remains as to the sign of the intrachain coupling $J_2$: magnetic susceptibility data can be fitted equally well by the FM or AF $J_2$ model~\cite{miura_spin_2006}, and this ambiguity is not resolved by contradictory DFT results that support either AF~\cite{xu_synthesis_2005, derakhshan_electronic_2007} or FM~\cite{koo_analysis_2008, schmitt_microscopic_2014} $J_2$ couplings. The magnitude of $J_3$ also remains to be determined, as comparable $J_2$ and $J_3$ couplings might invalidate a spin chain scenario.

To establish whether Na$_2$Cu$_2$TeO$_6$ represents a rare realization of the alternating AF-FM spin chains, here we perform inelastic neutron scattering (INS) experiments on a single crystal sample of Na$_2$Cu$_2$TeO$_6$ to determine the exchange coupling strengths. We confirm the spin gap originates from the dominant AF coupling $J_1$ (22.8~meV), and the interchain coupling $J_3$ is found to be much weaker (1.3~meV). Most importantly, we reveal the intrachain coupling $J_2$ to be FM with a strength ($-8.7$~meV) that is much higher than $J_3$, thus establishing Na$_2$Cu$_2$TeO$_6$ as a weakly-coupled alternating AF-FM chain compound. Through the DFT calculations, the alternating signs of the interchain couplings can be attributed to their different exchange paths. 

\textit{Experimental Details.} Polycrystalline Na$_2$Cu$_2$TeO$_6$ was utilized as a source material for crystal growth in a flux based on TeO$_2$ and Na$_2$CO$_3$.  Details for the polycrystal synthesis and characterization can be found in the Supplemental Materials~\cite{supp}. Powders of these materials in a ratio of 2(Na$_2$Cu$_2$TeO$_6$):1(Na$_2$CO$_3$):4(TeO$_2$) were mixed and loaded into a Pt crucible that was covered with a Pt lid.  The crucible was heated rapidly to 900~$^\circ$C where the melt was homogenized for 12~h in air.  The furnace was then cooled at 2~$^\circ$C/h to 500~$^\circ$C at which point it was turned off to cool.  The translucent green crystals were recovered by boiling the product-filled crucible in a hot aqueous solution of potassium hydroxide, followed by additional rinsing in deionized water.

INS experiments on Na$_2$Cu$_2$TeO$_6$ were performed on the fine-resolution Fermi chopper spectrometer SEQUOIA at the SNS of the ORNL. A single crystal (mass $\sim$140~mg) was aligned with the (001) vector vertical. A closed cycle refrigerator (CCR) was employed to reach temperatures $T$ down to 5~K. The incident neutron energy was $E_i  = 60$~meV, and the Fermi chopper frequency of 180~Hz and 420~Hz was selected in the high-intensity and high-resolution modes, respectively. Data were acquired by rotating the sample in 1$^{\circ}$ steps, covering a total range of 200$^{\circ}$ at 5 K and 100$^{\circ}$ at higher temperatures. Data reductions and projections were performed using MANTID~\cite{arnold_mantid_2014} and HORACE softwares~\cite{ewings_horace_2016}. 

%---------------------------------------------------------------------
\begin{figure}[t!]
\includegraphics[width=0.35\textwidth]{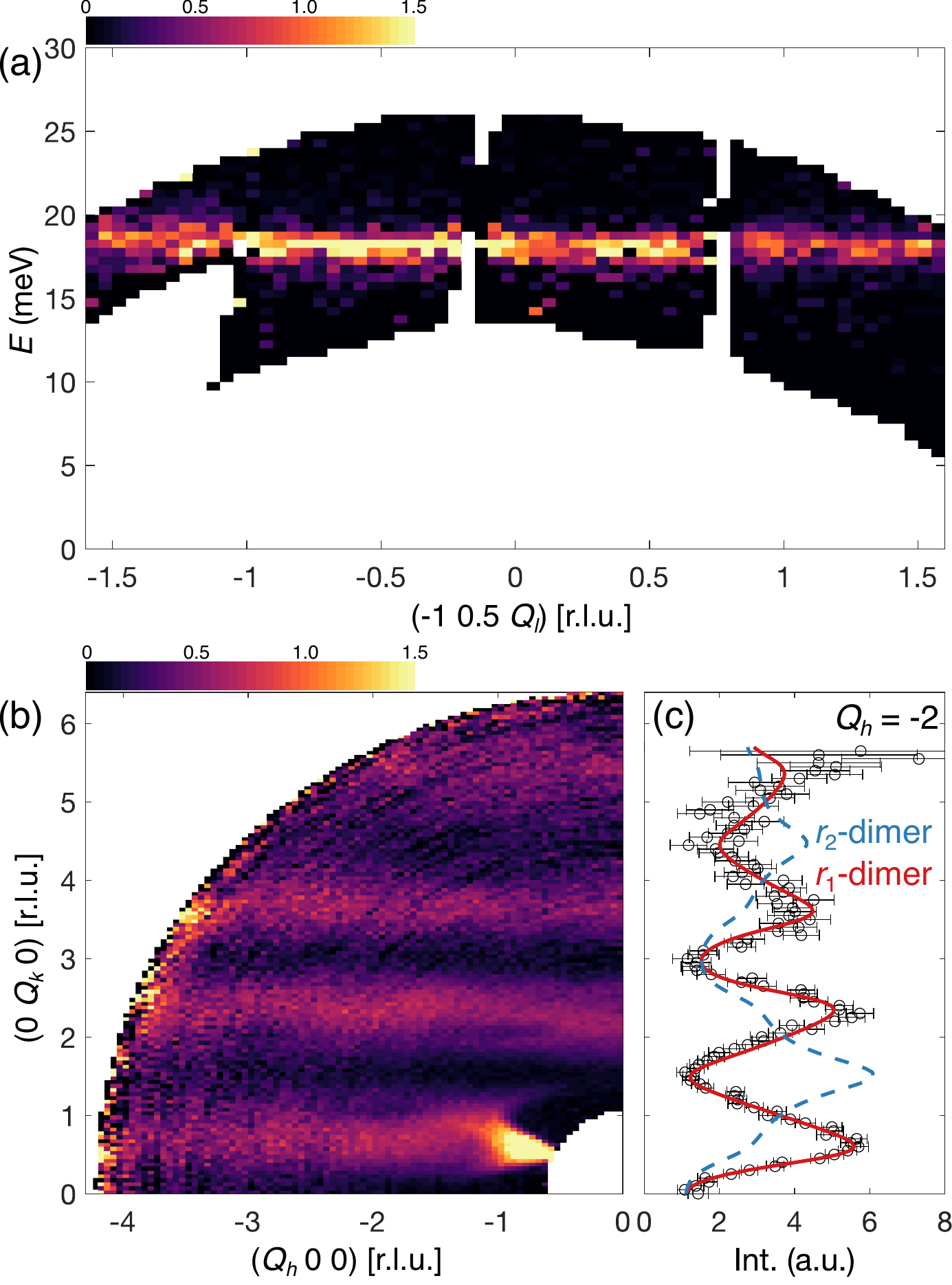}
\caption{(a) Scattering intensity of Na$_2$Cu$_2$TeO$_6$ measured along $(-1\ 0.5\ Q_l)$  at $T=5$~K with integration widths of $\Delta Q_h=0.2$ and $\Delta Q_k=0.1$. Measurements were performed in the high-resolution configuration. (b) Intensity map in the $(Q_h\ Q_k\ 0)$ plane integrated over ranges of $[17,\ 28]$~meV in $E$ and $[-2,\ 2]$ in $Q_l$. Measurements were performed in the high-intensity setup at $T=5$~K. (c) Scattering intensity along $Q_k$ integrated from the map in panel (b) over the range of $[-2.1,\ 1.9]$ in $Q_h$. The intensity can be described by the structure factor of the $r_1$-dimer model (red solid line) instead of the $r_2$-dimer model (blue dashed line), where $r_1\approx 2b/3$ and $r_2\approx b/3$ are the bond distances for $J_1$ and $J_2$, respectively.
\label{fig:lslice}}
\end{figure}
%---------------------------------------------------------------------

%---------------------------------------------------------------------
\begin{figure*}[t!]
\includegraphics[width=0.95\textwidth]{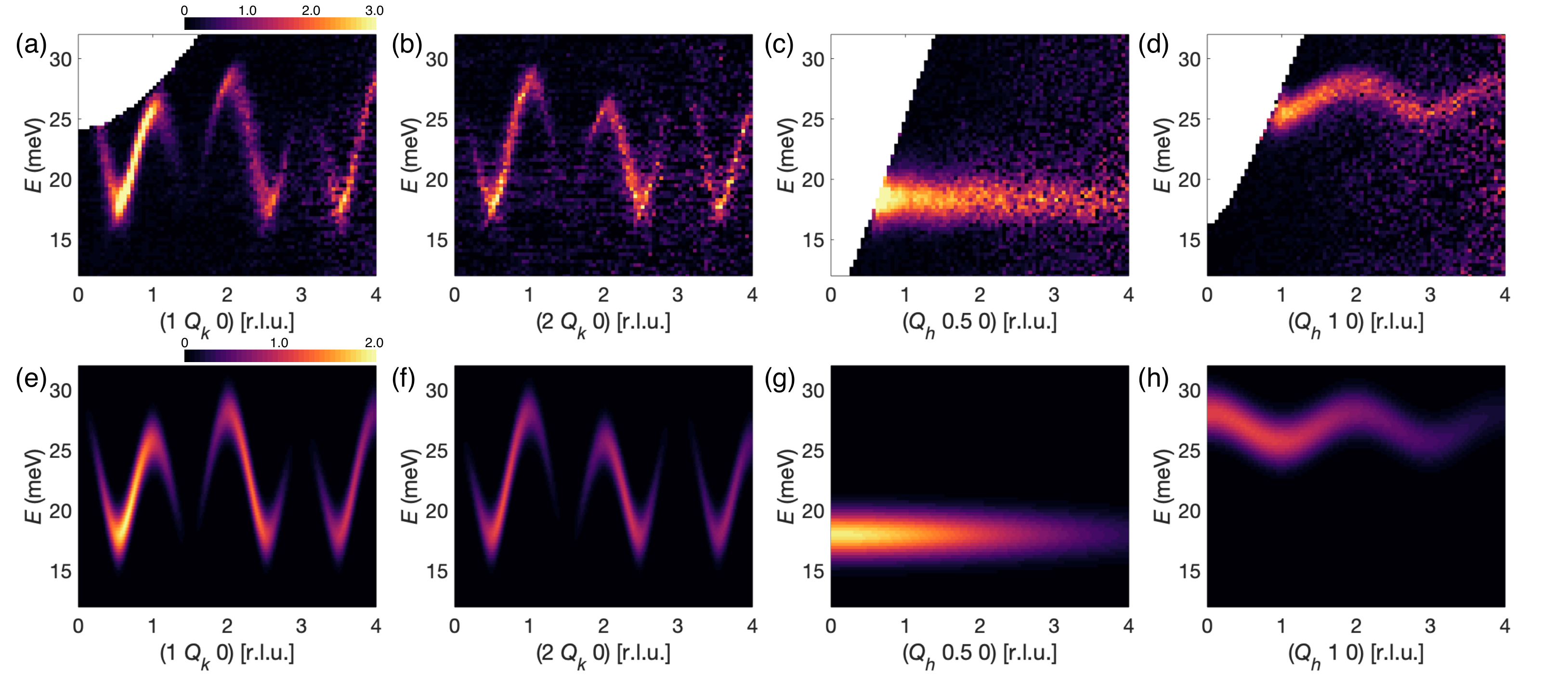}
\caption{(a-d) Intensity maps of Na$_2$Cu$_2$TeO$_6$ measured along (1 $Q_k$ 0), (2 $Q_k$ 0), ($Q_h$ 0.5 0), and ($Q_h$ 1 0) at $T = 5$~K using the high-intensity configuration. The integration widths are $\Delta Q_h=0.1$, $\Delta Q_k=0.1$, and $\Delta Q_l=2$. (e-h) Intensity maps calculated through the random phase approximation with fitted coupling strengths $J_1=22.78(2)$ meV, $J_2=-8.73(4)$~meV, and $J_3=1.34(3)$~meV. An instrumental energy resolution of 3.0 meV was convolved with the calculated spectrum. Panels (a)-(d) and (e)-(h) are shown on the identical intensity scale, respectively.
\label{fig:sma}}
\end{figure*}
%---------------------------------------------------------------------
First-principles calculations using the projector augmented wave (PAW) method were performed based on the DFT as implemented in the Vienna {\it ab initio} Simulation Package (VASP) code \cite{kresse_efficient_1996, kresse_from_1999, blochl_projector_1994}. The generalized gradient approximation (GGA) and the revised Perdew-Burke-Ernzerhof (PBEsol) function were used to treat the electron exchange-correlation potential \cite{perdew_generalized_1996, perdew_restoring_2008}. Based on the {\it ab initio} ground-state wave function, the Cu-site centered Wannier functions (WFs) with orbital $d_{x^2-y^2}$ were constructed using the WANNIER90 code \cite{marzari_maximally_1997, mostofi_wannier_2008}.

\textit{Results.} As shown in Fig.~\ref{fig:xrd}(a), the large separation of $c=5.92$~\AA\ between the honeycomb layers indicates likely negligible interlayer couplings. This is immediately confirmed in our single-crystal INS experiment. Figure~\ref{fig:lslice}(a) plots a representative slice of the measured spectrum along the $Q_l$ direction that is perpendicular to the $ab$ plane. A single excitation mode is observed at $\sim18$~meV, which is flat within the instrument resolution (1.6 meV at the elastic line). In contrast,  strong dispersion in the energy range of $E=$ [17, 28]~meV are observed in the $ab$ plane as summarized in Fig.~\ref{fig:sma}(a-d), thus confirming all the related couplings to be within the honeycomb layers.

The lack of dispersion out of the $ab$ plane allows us to integrate data along $Q_l$ for better statistics. Fig.~\ref{fig:lslice}b plots the scattering intensity integrated in the range of $Q_l$ in [-2, 2] (r.l.u.) and $E$ in [17, 28] meV. Along the $Q_k$ direction, intensity is strongly modulated with a periodicity of $\sim1.5$~(r.l.u.). For dimer systems, it is established that intensity of the triplon excitations is modulated by the dimer structure factor $S(\bm{Q})\propto [1-\cos(\bm{Q}\cdot \bm{r})]$, where $\bm{r}$ is the vector that connects the two spin sites within the dimer~\cite{leuenberger_spin_1984, xu_triplet_2000, cavadini_magnetic_2001}. Therefore, the modulation along $Q_k$ indicates that dimers in Na$_2$Cu$_2$TeO$_6$ are forming along the $b$ axis, and its periodicity tells the bond distance $r$ within the dimers. As shown in Fig.~\ref{fig:lslice}(c), the model that assumes dimers forming over the $J_1$ bonds with a distance of $r_1=2b/3$ accurately describes the intensity modulation, thus confirming the $J_1$ couplings to be AF and dominant as proposed by the DFT calculations~\cite{miura_spin_2006, derakhshan_electronic_2007, koo_analysis_2008, schmitt_microscopic_2014}.

Figure~\ref{fig:sma}(a-d) summarizes the dispersion along $Q_h$ and $Q_k$ in the $ab$ plane. Along the chain direction ($Q_k$), the triplon band reaches its lowest energy of $\sim18$~meV at $Q_k = N+1/2$ with integer $N$, indicating the FM character of the interdimer $J_2$ couplings~\cite{tennant_neutron_2003, stone_quantum_2007}. Different from the conventional chain compounds, the top of the dispersion varies at successive integer $Q_k$ positions, which might arise from the interchain couplings. As compared in Fig.~\ref{fig:sma}(c) and (d), although the triplon band looks flat along ($Q_h$ 0.5 0) at the bottom of the band, dispersion with a bandwidth of $\sim3$~meV is observed along ($Q_h$ 1 0) at the top of the band, which suggests weak but non-negligible $J_3$ couplings.
%---------------------------------------------------------------------
\begin{figure}[t!]
\includegraphics[width=0.46\textwidth]{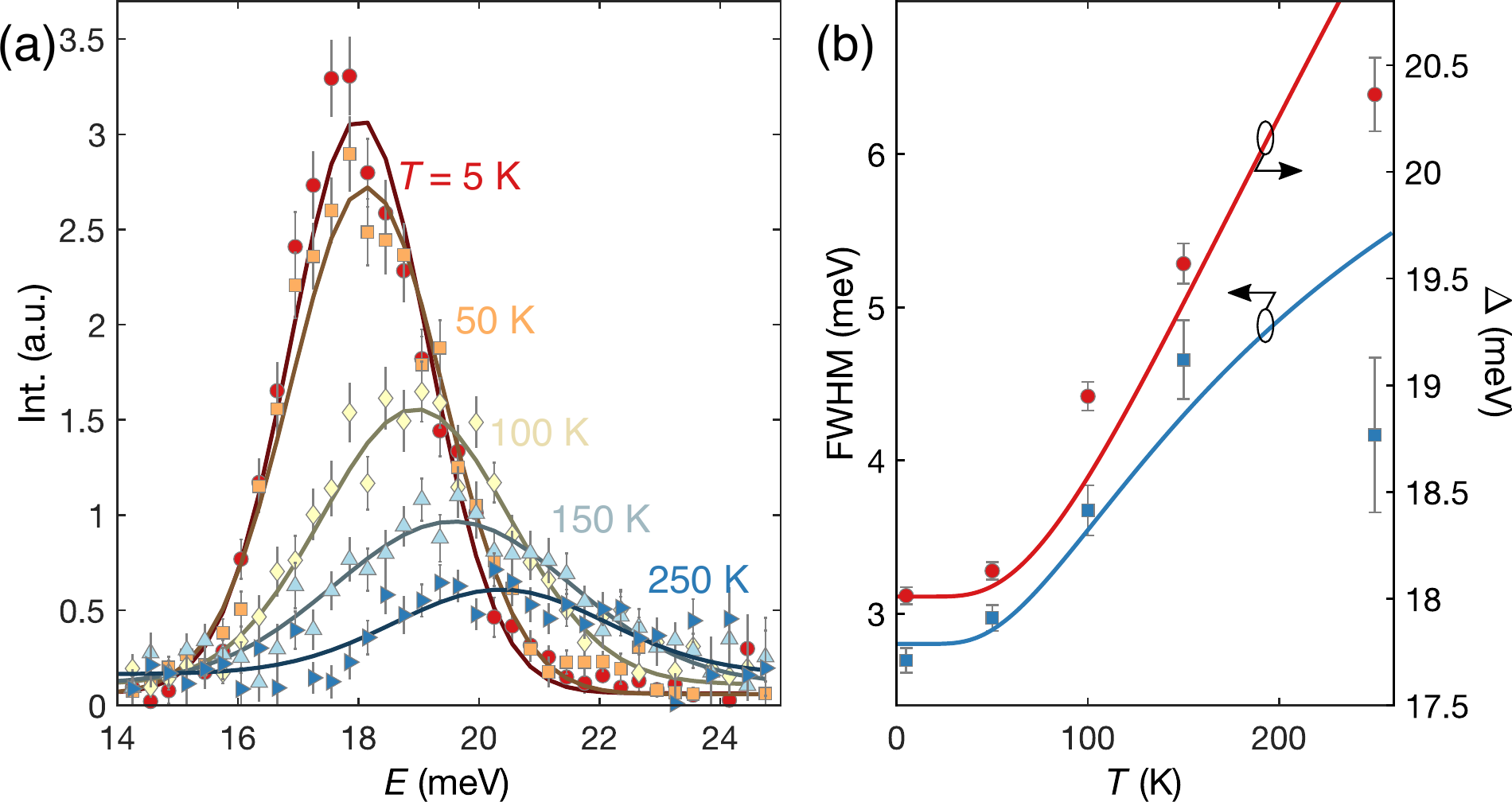}
\caption{(a) Energy dependence of the scattering intensity at (1 0.5 0) measured at $T=5$ (circles), 50 (squares), 100 (diamonds), 150 (up-pointing triangles), and 250 (right-pointing triangles) K. Solid lines are gaussian fits. (b) Temperature dependence of the fitted gap size  $\Delta$ (circles, righ axis) and full width at half maximum FWHM  (squares, left axis). Solid lines are fits using the empirical formula in Eq.~(\ref{eqn:gap}). The fitted parameters are $\alpha = 2.3(8)$~meV, $\gamma = 6(2)$~meV, and $\Gamma_0 = 2.8(2)$~meV. An instrumental energy resolution of 3 meV was convolved with the calculated spectrum.
\label{fig:tdep}}
\end{figure}
%---------------------------------------------------------------------

The triplon dispersion in dimer systems can often be analyzed through the random phase approximation~\cite{haley_standard_1972, leuenberger_spin_1984, stone_singlet_2008, stone_dispersive_2008, allenspach_multiple_2020}. Under this approximation, the dispersion relation can be written as
\begin{equation}
\hbar\omega(\bm{Q}) = \sqrt{J_1^2 +J_1\mathcal{J}(\bm{Q})R(T)}\rm{,}
\label{eqn:disp}
\end{equation}
where $R(T)$ describes the population difference between the singlet and triplet states~\cite{leuenberger_spin_1984, stone_singlet_2008} and for Na$_2$Cu$_2$TeO$_6$ can be approximated by 1 at $T = 5$ K due to the large excitation gap. $\mathcal{J}(\bm{Q})$ is the Fourier sum of interactions beyond dimer exchange:
\begin{equation}
\mathcal{J}(\bm{Q})=-J_2\cos(2\pi Q_k) - 2J_3\cos(\pi Q_h)\cos(\pi Q_k)\rm{.}
\end{equation}
Experimental dispersion values were extracted from Gaussian fits to constant $Q$ scans at 140 points throughout the measured reciprocal space volume, which were then fitted by the dispersion relation in Eq.~(\ref{eqn:disp}). The fitted coupling strengths are $J_1=22.78(2)$ meV, $J_2=-8.73(4)$~meV, and $J_3=1.34(3)$~meV. The calculated spectra are summarized in Fig.~\ref{fig:sma}(e-h) for comparison with the experimental data. 
The FM character of the $J_2$ couplings is thus unambiguously established, and the weakness of the interchain $J_3$ justifies the description of Na$_2$Cu$_2$TeO$_6$ as a weakly-coupled alternating AF-FM chain compound.

%---------------------------------------------------------------------
\begin{figure}[t!]
\includegraphics[width=0.35\textwidth]{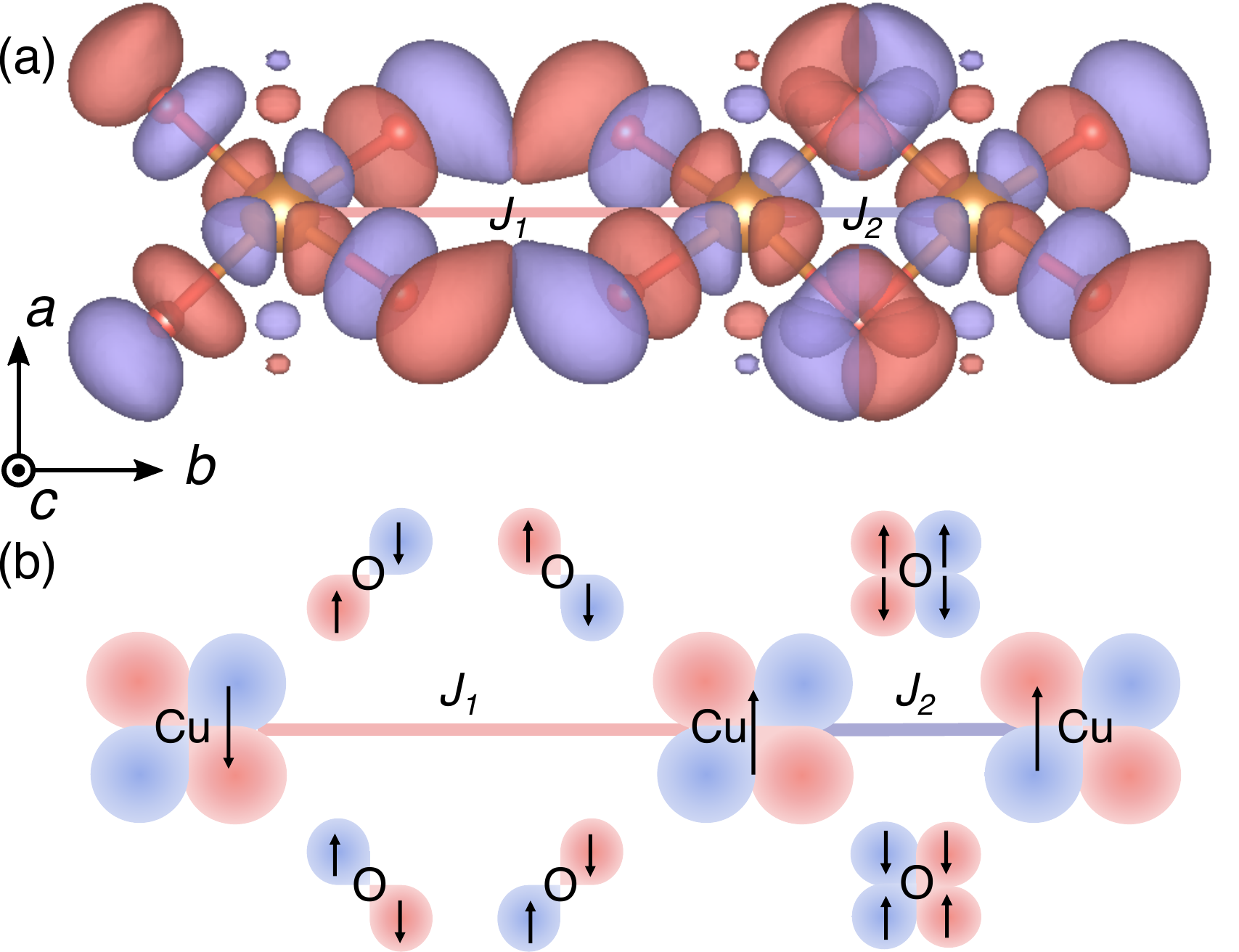}
\caption{(a) Wannier functions of the Cu $3d_{x^2-y^2}$ orbital, an antibonding combination of the Cu $3d_{x^2-y^2}$  and O $2p$ states, viewed along the $c$ axis. Different colors represent the +/$-$ signs of the Wannier function. (b) Diagrams for the super-superexchange and superexchange couplings between the nearest-neighbor Cu $3d_{x^2-y^2}$ orbitals via oxygen $2p$ ligands. For the $J_1$ path, the Cu-O$\cdots$O-Cu super-superexchange leads to the AF alignment of the nearest-neighbor Cu ions. For the $J_2$ path, Cu-O-Cu superexchange with a bonding angle of $90^{\circ}$ results in a FM exchange between the nearest-neighbor Cu ions.
\label{fig:wannier}}
\end{figure}
%---------------------------------------------------------------------

In Haldane chains the existence of a spin gap is known to protect the spin entanglement~\cite{wierschem_characterizing_2014}. Na$_2$Cu$_2$TeO$_6$ can serve as a model system to test that the same ideas are valid for alternating spin chains. In long-range ordered magnets, the excitation gap often decreases at higher $T$ due to reduced ordering moments. As a contrast, the gap of Haldane chains increases with $T$ due to the reduction in the coherent length and the consequent finite-size effect~\cite{jolicur_sigma_1994, nightingale_gap_1986}. The $E$ scans at $\bm{Q}=$ (1, 0.5, 0) shown in Fig.~\ref{fig:tdep}(a) indeed reveal an increased gap at elevated $T$. We parameterize the lineshape using a Gaussian function that is independently fit at each temperature, and the value of $\Delta$ is the centroid of the Gaussian peak. As summarized in Fig.~\ref{fig:tdep}(b), in a large range below $\sim150$~K, the fitted gap size $\Delta$ and full width at half maximum (FWHM) $\Gamma$ exhibit an activated behavior that are characteristic of Haldane chains~\cite{zheludev_spin_1996, bera_critical_2015}:
\begin{align}
\Delta(T) &= \Delta_0+ \sqrt{\alpha T}\exp(-\Delta_0/T)\nonumber \\
\Gamma(T) &= \Gamma_0 + \gamma\exp(-\Delta_0/T)\rm{,}
\label{eqn:gap}
\end{align}
where $\alpha$, $\gamma$, and $\Gamma_0$ are fitting parameters, $\Delta_0$ is the gap size at 0 K and is fixed at 18.0 meV. The validity of the activated behavior further confirms the similarity between the $S=1/2$ alternating AF-FM chains and the integer-spin Haldane chains, thus revealing the robustness of the topological ground state against weak interchain couplings.

\textit{Discussion.} The emergence of spin chains in Na$_2$Cu$_2$TeO$_6$ can be ascribed to the distortion of the honeycomb lattice. As shown in Fig.~\ref{fig:xrd}(b), the exchange paths of $J_3$ involve the longest Cu-O bonds b3 of the distorted octahedra. Therefore, the $J_3$ couplings are expected to be weak as the unoccupied Cu $3d_{x^2-y^2}$ orbitals disfavor the elongated bond direction, which reduces the electron hopping between the chains.

The sign of the interchain couplings $J_2$ is more subtle, and different scenarios exist in the previous DFT calculations~\cite{xu_synthesis_2005, derakhshan_electronic_2007, koo_analysis_2008, schmitt_microscopic_2014}. As summarized in the Supplemental Materials, our DFT calculations confirm the alternating FM and AF intrachain couplings in agreement with the experiment.  The contrasting $J_1$ and $J_2$ couplings can be understood through the Wannier functions (WFs) as plotted in Fig.~\ref{fig:wannier}(a). Due to the contributions from the O $2p$ states, the WFs overlap directly over the $J_1$ paths but are almost orthogonal over the $J_2$ path. Therefore, the $J_1$ path, in spite of its longer distance, develops a stronger coupling than that over the $J_2$ path. Based on the WFs, the signs of the couplings can be understood through the Goodenough-Kanamori-Anderson rules~\cite{anderson_anti_1950, goodenough_theory_1955, goodenough_an_1958, moskvin_angular_1975} as shown in Fig.~\ref{fig:wannier}(b). For the $J_1$ couplings, the Cu-O$\cdots$O-Cu super-superexchange leads to an AF interaction between the Cu$^{2+}$ spins. While for the $J_2$ couplings, the interaction become FM as the angle of \angle Cu-O-Cu  is close to $\sim 90^{\circ}$, which means a pair of orthogonal O~$2p$ orbitals with parallel spins are involved in the virtual electron hopping.

\textit{Conclusions.} Neutron scattering experiments have been performed on the honeycomb-lattice compound Na$_2$Cu$_2$TeO$_6$ to study its spin correlations. A triplon excitation mode was observed, which exhibits strong dispersion along the chain but weak dispersion perpendicular to the chain. Under the random phase approximation, the intrachain couplings were found to be alternating AF and FM, and a weak interchain coupling was also established. The emergence of spin chains in Na$_2$Cu$_2$TeO$_6$ was ascribed to the distortion of the honeycomb lattice, and the alternating intrachain couplings were understood through the DFT calculations. Our works establish the existence of weakly-coupled alternating AF-FM spin-1/2 chains in Na$_2$Cu$_2$TeO$_6$ and reveal a robust gapped ground state that is similar to that of the integer-spin Haldane chains.

\textit{Acknowledgments.} We acknowledge helpful discussions with Jyong-Hao Chen, Tong Chen, and Tao Hong. This work was supported by the U.S. Department of Energy, Office of Science, Basic Energy Sciences, Materials Sciences and Engineering Division. This research used resources at the Spallation Neutron Source (SNS), a DOE Office of Science User Facility operated by the Oak Ridge National Laboratory (ORNL).

\clearpage
\newpage

\renewcommand{\thefigure}{S\arabic{figure}}
\renewcommand{\thetable}{S\arabic{table}}

\renewcommand{\theequation}{\arabic{equation}}

\makeatletter
\renewcommand*{\citenumfont}[1]{S#1}
\renewcommand*{\bibnumfmt}[1]{[S#1]}
\def\clearfmfn{\let\@FMN@list\@empty}  
\makeatother
\clearfmfn

\setcounter{figure}{0} 
\setcounter{table}{0}
\setcounter{equation}{0} 

\onecolumngrid
\begin{center} {\bf \large Weakly-coupled alternating $S=1/2$ chains in the distorted honeycomb-lattice compound Na$_2$Cu$_2$TeO$_6$ \\
 Supplementary Information} \end{center}
 \maketitle

\vspace{0.5cm}
%\twocolumngrid
\section{Polycrystal synthesis and characterization}
Polycrystalline samples of Na$_2$Cu$_2$TeO$_6$ were prepared using a solid-state method. Powders of Na$_2$CO$_3$ (99.997$\%$, Alfa Aesar), TeO$_2$ (99.99$\%$, Beantown Chemical) and CuO (99.995$\%$, Alfa Aesar) were mixed in a molar ratio of 1.03:1:2. The mixed powders were ground thoroughly in air and pressed into pellets with $\frac{1}{2}''$ diameter.  A furnace was preheated to 660~$^\circ$C before the pellets were loaded in the furnace to minimize the loss of Na$_2$CO$_3$. The temperature was held at 660~$^\circ$C for 8 hours before cooling down to room temperature.

Neutron diffraction experiments on a powder sample of Na$_2$Cu$_2$TeO$_6$ were performed on the time-of-flight  powder diffractometer POWGEN at the Spallation Neutron Source (SNS) of the Oak Ridge National Laboratory (ORNL). The 	powder (mass $\sim$8.4~g) was placed in a vanadium can.	Data were acquired at 300 K and 20 K using the 1.5~\AA~instrumental configuration with an incoming neutron wavelength band of 0.967$-$2.033~\AA. Rietveld refinements of the neutron diffraction data were performed using  the FULLPROF program~\cite{rodriguez_recent_1993}.

\begin{figure}[b]
\centering
\includegraphics[width=0.7\textwidth]{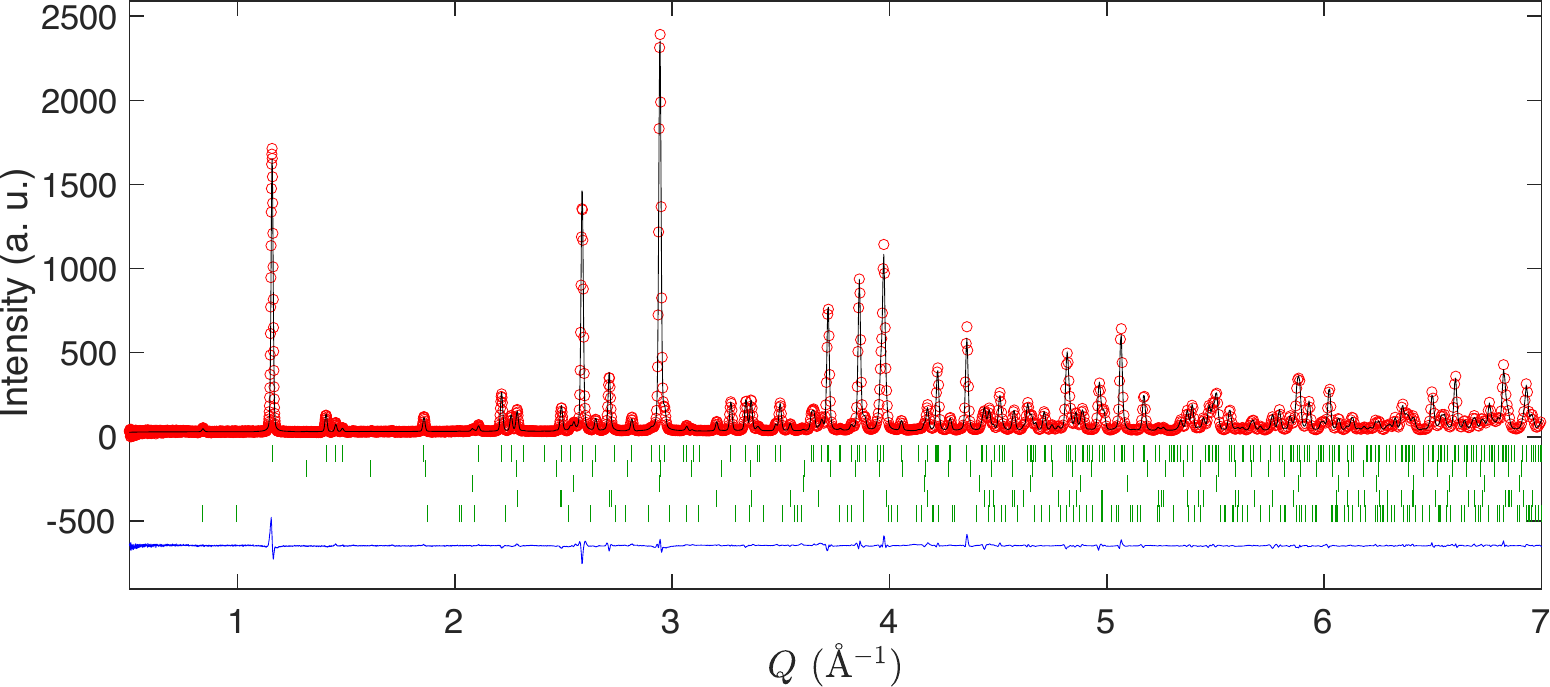}
\caption{Refinement result of the powder neutron diffraction data measured at $T = 20$~K. Data points are shown as red circles. The calculated pattern is shown as the black solid line. The vertical bars from top to bottom indicate the positions of the structural Bragg peaks for Na$_2$Cu$_2$TeO$_6$, Cu$_3$TeO$_6$, Cu$_2$O, CuO, and the magnetic Bragg peaks for CuO. The blue line at the bottom shows the difference of measured and calculated intensities.}
\label{figs:refine}
\end{figure}

Figure~\ref{figs:refine} summarizes the refinement results for the powder neutron diffraction data collected at $T = 20$~K, which is far below the $\sim127$~K gap revealed in magnetic susceptibility. The Na$_2$Cu$_2$TeO$_6$ phase (fractional weight $\sim88$\%) can be refined using the monoclinic $C2/m$ space group with lattice parameters $a = 5.6951(1)$~\AA, $b = 8.6560(1)$~\AA, $c=5.9236(1)$~\AA, and $\beta =113.741(1)^\circ$, which are consistent with the previous report~\cite{xu_synthesis_2005s}. About $9\%$ deficiency was observed at the Na sites. The Cu ions occupy the $4g$ sites at [0, 0.665(1), 0], leading to distances of $d_1=5.806(3)$~\AA$~\approx b/3$ and $d_2=2.850(3)$~\AA~$\approx 2b/3$~for the $J_1$ and $J_2$ bonds, respectively.  Minority phases of Cu$_3$TeO$_6$ ($\sim 0.3$~\%), Cu$_2$O ($\sim0.9$~\%), and CuO ($\sim9.8$~\%) were detected in the powder sample. The magnetic phase of CuO was also included in the refinement~\cite{wang_magnetoelectric_2016}, which contributes to the weak reflection at $Q\sim 0.84$~\AA$^{-1}$. The overall goodness-of-fit parameters are $R_p=10.8\%$, $R_{wp}=7.2\%$, and $\chi^2=24.9$. Most importantly, no magnetic reflection of Na$_2$Cu$_2$TeO$_6$ can be discerned in our diffraction data at $T = 20$~K, which confirms the quantum disordered ground state of the gapped spin system~\cite{affleck_rigorous_1987}.

\section{DFT calculations}
\begin{figure}
\centering
\includegraphics[width=0.48\textwidth]{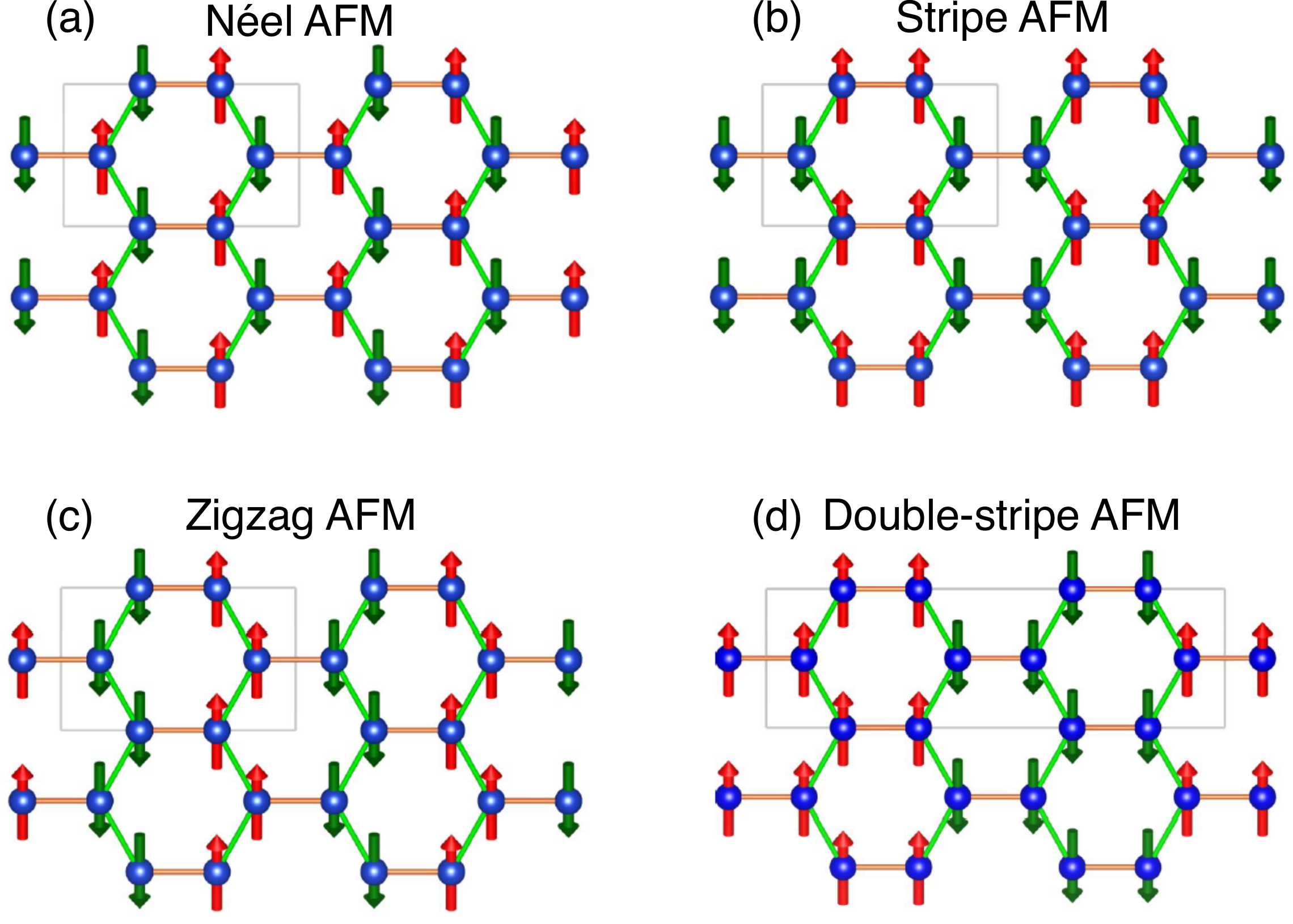}
\caption{Antiferromagnetic configurations considered in the DFT analysis. (a) N\'{e}el AFM (N-AFM), (b) stripe-AFM (S-AFM), (c) zigzag-AFM (Z-AFM), and (d) double-stripe AFM (D-AFM). The gray rectangle represents the minimum magnetic unit cell for each configuration. To eliminate the effect of choosing different cell sizes, the unit cell as shown in (d) is adopted in our calculations to compute energies for different magnetic configurations.}
\label{magnetism}
\end{figure}

% Spin chains are one of the simplest examples that demonstrate the determinant role of quantum fluctuations in magnetic systems. For spins $S=1/2$, the one-dimensional (1-D) character of the chain leads to fractional spinon excitations and critical correlations that are described as Tomonaga-Luttinger liquids, which contrasts the exponential correlations and gapped triplons excitations for $S=1$. 
To better understand the magnetic properties of Na$_2$Cu$_2$TeO$_6$, we introduced the electron correlation by using GGA plus $U_{\rm eff}$ ($U_{\rm eff} = U-J$) with the Dudarev format on the Cu site \cite{dudarev_electron_1998}. As shown in Fig.~\ref{magnetism}, we considered four different antiferromagnetic (AF) configurations together with the ferromagnetic (FM) state.

Using the experimental crystal structure, we calculated the energies of five magnetic configurations as a function of $U_{\rm eff}=0$, 2, 4, 6, and 8 eV  in Table~\ref{table1}. The exchange interactions can be estimated by mapping the calculated total energies for each magnetic state to the classical Heisenberg model. Four AFM equations are used to calculate four unknown parameters and the extracted spin exchange parameters can be found in Table~\ref{table2}.

\begin{table}[b!]
\centering
\caption{List of energy equations and calculated energies of the five collinear spin configurations used to estimate the magnetic exchange integrals, where $S=\frac{1}{2}$ is the magnetic moment. The double-stripe AFM state is taken as the energy reference. All energies in units of meV.}
\begin{tabular*}{0.98\textwidth}{@{\extracolsep{\fill}}lcccccc}
\hline
\hline
		\multirow{2}*{Confg.} &\multirow{2}*{Energy equations} & \multicolumn{5}{c}{Energy (meV)} \\
		\cline{3-7}
		~ & ~ & $U_{\rm eff}$ (eV)= 0 & 2  & 4  &6  &8  \\
\hline
FM   & $E_0+4J_1S^2+4J_2S^2+8J_3S^2$ & 269& 172& 116& 77& 49\\
N-AFM & $E_0-4J_1S^2-4J_2S^2-8J_3S^2$ & 38& 33& 28& 22& 16\\
S-AFM & $E_0+4J_1S^2+4J_2S^2-8J_3S^2$ & 252& 160& 107& 71& 44\\
Z-AFM & $E_0-4J_1S^2-4J_2S^2+8J_3S^2$& 55& 46& 38& 29& 21\\
D-AFM & $E_0-4J_1S^2+4J_2S^2$ & 0& 0& 0& 0& 0\\
\hline
\hline
\end{tabular*}
\label{table1}
\end{table}

\begin{table}[b!]
\centering
\caption{List of magnetic exchange couplings obtained in our DFT calculations, in units of meV. }
\begin{tabular*}{0.48\textwidth}{@{\extracolsep{\fill}}ccccccc}
\hline
\hline
Coupling&$U_{\rm eff}$ (eV) = 0 & 2  & 4 & 6  & 8 \\
\hline
$J_1$   & 130.3 & 83.3& 56.0& 37.3& 23.3\\
$J_2$  &-23.3 & -19.8& -16.3& -12.8& -9.3\\
$J_3$ & 4.3 & 3.3& 2.5& 1.8& 1.3\\
\hline
\hline
\end{tabular*}
\label{table2}
\end{table}

For all of the values of $U_{\rm eff}$ employed here, the super-superexchange coupling $J_1$ is AF, and its coupling strength is several times higher than that of the FM coupling $J_2$. The  $J_3$ coupling is AF and is much weaker in strength than $J_1$ and $J_2$. Thus, the two strongest spin exchange couplings $J_1$ and $J_2$ form alternating chains, which is consistent with the experimental results in the main text. With $U_{\rm eff}$ in the range of [0, 8] eV, the ratio of $J_2/J_1$ has maximum value of $-0.399$ and a minimum value of $-0.179$, covering the experimentally determined value of $-0.383(2)$.  The calculated value of $J_3/J_1$ has a maximum value of 0.056 and a minimum value of 0.033.  The experimentally determined ratio of $J_3/J_1$ was found to be 0.059(1), which is close to the upper limit of the calculated value.

All our present DFT results show good agreement with the previous theoretical works~\cite{koo_analysis_2008s, schmitt_microscopic_2014s}. As explained in Ref.~\cite{koo_analysis_2008s}, the DFT calculations in Ref.~\cite{xu_synthesis_2005s} neglects the ferromagnetic contributions to the exchange coupling, thus predicting the wrong sign for $J_2$ when the coupling strength is weak~\cite{whangbo_spin_2003s}.  While in Ref.~\cite{derakhshan_electronic_2007s}, although the authors emphasized that the $J_2$ coupling should be AF by comparing their total energy with different spin arrangements, not sufficient information about the details of the calculation, such as energies or spin arrangements used, were presented in that publication. By contrast, all the details of the energies and mapping equations for different magnetic configurations are listed clearly in the present work.


%\bibliography{NCTO_chain}
\end{document}